\documentclass[letterpaper,11pt]{article}

\usepackage{amsthm}
\usepackage{amsmath}
\usepackage{amsfonts}
\usepackage{graphicx}
\newtheorem{thm}{Theorem}[section]

\theoremstyle{definition}

\theoremstyle{remark}

\numberwithin{equation}{section}

\def\sign{\operatorname{sign}}

\title{A Lower Bound for Fourier Transform Computation in a Linear Model Over 2x2  Unitary Gates Using Matrix Entropy }
\author{Nir Ailon}

\begin{document}

\maketitle

\def\sign{\operatorname{sgn}}
\def\dim{n}
\def\C{\mathbb C}
\def\P{\mathcal P}
\def\trace{\operatorname{tr}}
\def\diag{\operatorname{diag}}
\def\rank{\operatorname{rank}}
\def\F{{\mathcal F}}
\def\Id{\operatorname{Id}}

\begin{abstract}
Obtaining a non-trivial (super-linear) lower bound for computation of the Fourier transform in the linear 
circuit model has been a long standing open problem.  All lower bounds so far have made
strong restrictions on the computational model.  One of the most well known results,
by Morgenstern from 1973, provides an $\Omega(n \log n)$ lower bound for the \emph{unnormalized} FFT  
when the constants used in the computation are bounded.  The proof uses a  potential function related
to a determinant.   The determinant of the unnormalized
Fourier transform is $n^{n/2}$, and thus by showing that it  can grow by at most a constant factor after each step
yields the result.
 This classic result, however, does not
explain why the \emph{normalized} Fourier transform, which has a unit determinant, should take $\Omega(n\log n)$ steps to compute.
In this work we show that in a layered linear circuit model restricted to unitary $2\times 2$ gates, one obtains an $\Omega(n\log n)$
lower bound.  The well known FFT works in this model.
The main argument concluded from this work is that
 a potential function that might eventually help  proving the $\Omega(n\log n)$ conjectured  lower bound for computation of Fourier transform  is not related
to matrix determinant, but  rather to a notion of matrix entropy.
\end{abstract}
\section{Introduction}
The Fast Fourier Transform \cite{CooleyT64} is a method for computing the complex Fourier transform of order $n$ 
in time $O(n\log n)$ using a so called linear algorithm.  A linear algorithm, as defined in
\cite{Morgenstern:1973:NLB:321752.321761},  is a sequence $\F_0,\F_1,\dots$, where each $\F_i$
is a set of affine functions,  for each  $i\geq 0$ $\F_{i+1} = \F_i \cup \{\lambda_i f + \mu_i g\}$ for some  $\lambda_i, \mu_i\in \C$ and $f,g\in \F_i$, and $\F_0$ contains (projections onto) the input variables as well as constants.

It is trivial  that computing the Fourier Transform requires a super-linear  number of steps, but no non-trivial lower bound is 
known.  
In 1973, Morgenstern proved that if the modulus of the $\lambda_i$'s and $\mu_i$'s is bounded by $1$
then the number steps required for computing the Fourier transform in the linear algorithm model is at least 
$\frac 1 2 n\log_2 n$.  It should be noted that Cooley and Tukey's FFT indeed can be expressed as a linear algorithm
with coefficients of the form $e^{ix}$ for some real $x$, namely, complex numbers of unit modulus.

The main idea of Morgenstern is to define a potential function for each $\F_i$ in the linear algorithm sequence,
equaling the maximal determinant of a square submatrix in a certain matrix corresponding to $\F_i$.
The technical step is to notice that the potential function can at most double in each step.
It should now be noted that Morgenstern's lower bound applies to the \emph{unnormalized} Fourier transform,
the determinant of which is $n^{n/2}$, hence the lower bound of $\frac 1 2 n\log_2 n$.

The determinant of any square sub-matrix of the  normalized  Fourier transform, however, is $1$.  
 Morgenstern's method can therefore not be used
to derive any useful lower bound for computing the normalized Fourier transform in the linear algorithm model
with constants of at most  unit modulus.  Using constants of modulus $1/\sqrt 2$ in the FFT,  on the other hand,
does compute the normalized Fourier transform in $O(n\log n)$ steps. 

The situation is quite odd.  The normalized and unnormalized Fourier transforms are proportional to each other, hence
there shouldn't be a real big difference between their computational complexities.  Can we obtain a meaningful lower bound
for computing the normalized Fourier transform?   In this work we show that such a bound is possible using a further
restriction of the computational model considered by Morgensten.  The main point is that for lower bounding
the computational cost of the normalized Fourier transform we should not be looking at determinants, but rather
at a certain type of \emph{entropy} related to the matrices.

\section{The Unitary Layered Circuit}
Our model of computation consists of layers $L_0,\dots, L_m$, each  containing exactly $n$ nodes and representing
a vector in $\C^n$. The first layer, $L_0 \in \C^n$, is the input.  The last layer $L_m\in \C^n$ is the output.
For each layer $i\geq 1$ there are two  indices $k_i, \ell_i \in [n]$, $k_i < \ell_i$, and a complex unitary matrix $$ A_i = \left ( \begin{matrix} a_i(1,1) & a_i(1,2) \\ a_i(2,1) & a_i(2,2) \end{matrix} \right )\ .$$
For each $j \not \in\{k_i, \ell_i\}$, $L_i(j) = L_{i-1}(j)$.  The values of $L_i(k_i)$ and $L_i(\ell_i)$ are given as
$$ \left (\begin{matrix} L_i(k_i) \\ L_i(\ell_i) \end{matrix} \right ) = A_i  \left (\begin{matrix} L_{i-1}(k_i) \\ L_{i-1}(\ell_i) \end{matrix} \right )\ .$$

In words, the next layer is obtained from the current layer by applying a $2$-by-$2$ unitary transformation on two coordinates.
Compared to Morgenstern's model of computation the unitary layered circuit is strictly weaker.  To see why it is not stronger, notice that the matrix elements of $A_i$ all have modulus at most $1$.  It is strictly weaker because it uses only unitary transformations, but also because it has a bounded
memory of $n$ numbers at any given moment.  Indeed, it is not possible in layer $L_{i+1}$ to use a coordinate of $L_{i'}$
for $i' < i$.  Still, the normalized FFT is implemented as a unitary layered circuit with $m=O(n\log n)$.

\begin{thm}
If a layered circuit given by $A_1,\dots, A_m \in \C^{2\times 2}$, $k_1,\dots, k_m\in[n]$ and $\ell_1,\dots, \ell_m\in[n]$ computes the normalized 
Fourier transform, then $m\geq \frac 1 2 n\log_2 n$.
\end{thm}
\begin{proof}
For a matrix $M$ and a set $I\subseteq [n]$ of indices, let $M[I]$ denote the principal minor corresponding to the set $I$.
For $i=1,\dots,m$ let $\tilde A_i$ denote the matrix defined so that $\tilde A_i[\{k_i,\ell_i\}] =  A_i$, $\tilde A_i[[n]\setminus\{k_i,\ell_i\}] = \operatorname{Id}$ and $\tilde A_i(p,q) = 0$ whenever exactly one of $p,q$ is in $\{k_i,\ell_i\}$.
It is clear that
$$ L_i = \tilde A_i \tilde A_{i-1} \cdots \tilde A_1 L_0\ .$$
It hence makes sense to define $M_i = \tilde A_i \tilde A_{i-1} \cdots \tilde A_1$.  Note that $M_m = F$, where $F$ is the normalized
FFT matrix.  We also define $M_0 = \Id$.
For a matrix $M$, we now define a potential function
$$ \Phi(M) = -\sum_{p,q} |M(p,q)|^2\log |M(p.q)|^2\ ,$$
where we formally define $0\log 0$ to be $\lim_{x\rightarrow 0^+} x\log x = 0$.
For a unitary matrix $M$, we notice that $\Phi(M)$ is the sum
of the Shannon entropies of the probability vectors
given by the squared moduli of the elements of each row.
Also notice that $\Phi(\Id) = 0$ and $\Phi(F) = n\log_2 n$, as all elements have modulus $1/\sqrt n$.
We now show that for any $i\geq 1$,
\begin{equation}\label{key} \Phi(M_i) - \Phi(M_{i-1})\leq 2\ .\end{equation}
This clearly implies the theorem statement.

To see (\ref{key}),  notice that $M_i$ is obtained from $M_{i-1}$ by replacing rows $k_i$ and $\ell_i$  as follows.  If $x$ and $y$ denote rows $k_i$ and $\ell_i$ of $M_{i-1}$, respectively, and $x',y'$ the corresponding rows of $M_i$ , then
 $x'= a_i(1,1) x + a_i(1,2) y$ and $y'=a_i(2,1)x + a_i(2,2) y$.  All other rows remain untouched.
We also have by orthonormality  that $\|x\|^2=\|y\|^2=\|x'\|^2=\|y'\|^2=1$ and that for all $j\in [n]$,
$$ |x'(j)|^2 + |y'(j)|^2 = |x(j)|^2 + |y(j)|^2 =: r(j) \ .$$

Now let $\P_r$ denote the set of pairs of vectors $(\alpha,\beta)\in [0,1]^n\times [0,1]^n$ 
satisfying:
\begin{enumerate}
\item $\sum_{j=1}^n \alpha(j) \leq 1$
\item $\sum_{j=1}^n \beta(j) \leq 1$
\item $\alpha(j) + \beta(j) = r(j)$ for $j\in [n]$.
\end{enumerate}
For $(\alpha, \beta)\in \P_r$ now let $$\Phi(\alpha,\beta) = -\sum \alpha(j)\log_2 \alpha(j) - \sum \beta(j)\log_2 \beta(j)$$
\noindent (abusing notation).  Then it suffices to show that
$$ \sup_{(\alpha,\beta)\in \P_r} \Phi(\alpha,\beta) - \inf_{(\alpha,\beta)\in \P_r} \Phi(\alpha,\beta) \leq 2\ .$$
Indeed, this can be seen by noting that the supremum is obtained for $(\alpha,\beta)$ satisfying $\alpha(j)=\beta(j)=r(j)/2$ for $j\in [n]$ (giving $\Phi(\alpha,\beta) = -\sum_{j=1}^n r(j)\log \frac{r(j)} 2$)
and the infimum is bounded below by the pair $(\alpha,\beta)$ satisfying
$\alpha(j) = r(j), \beta(j) = 0$ for  $j$ odd and $\alpha(j)=0, \beta(j)=r(j)$ for $j$ even,
giving $\Phi(\alpha,\beta) = \sum_{j=1}^n r(j)\log r(j)$. 
(Note that this pair may lie outside $\P_r$.)
 The difference is at most $\sum_{j=1}^n r(j) = 2$.
\noindent
By the above discussion, both vector pairs
 $$(\alpha, \beta) =  ((|x(1)|^2,\dots, |x(n)|^2), (|y(1)|^2, \dots, |y(n)|^2))$$
and  $$(\alpha', \beta') =  ((|x'(1)|^2,\dots, |x'(n)|^2), (|y'(1)|^2, \dots, |y'(n)|^2))$$ are in $\P_r$.
 This means that $\Phi(M_i) \leq \Phi(M_{i-1})+2$, as required.
\end{proof}
\bibliographystyle{plain}
\bibliography{low_bound_fft}
\end{document}